\documentclass[a4paper,fleqn,usenatbib]{mn2e}
\usepackage{newtxtext,newtxmath}

\usepackage[T1]{fontenc}
\usepackage{ae,aecompl}
\usepackage{footnote}

\usepackage{graphicx}	
\usepackage{amsmath}	
\usepackage{amssymb}	
\usepackage[caption=false]{subfig}

\newcommand{\ha}{H$\alpha$}

\def \st{\ifmmode{  {\mathrm{st}}}\else{${  {\mathrm{st}}}$}\fi}
\def \nd{\ifmmode{  {\mathrm{nd}}}\else{${  {\mathrm{nd}}}$}\fi}
\def \rd{\ifmmode{  {\mathrm{rd}}}\else{${  {\mathrm{rd}}}$}\fi}
\def \th{\ifmmode{  {\mathrm{th}}}\else{${  {\mathrm{th}}}$}\fi}

\newcommand{\hbeta}{\rm H$\beta$}

\newcommand{\fluxa}{$10^{-15}$ erg s$^{-1}$ cm$^{-2}$ arcsec$^{-2}$}

\newcommand{\sulfur}{[S~{\sc ii}]}
\newcommand{\sulfuri}{[S~{\sc iii}]}
\newcommand{\nitrogen}{[N~{\sc ii}]}
\newcommand{\oxygen}{[O~{\sc iii}]}

\newcommand{\OII}{[O{\sc ii}]}

\newcommand{\ArIII}{[Ar {\sc iii}]}

\title[New Galactic Planetary Nebulae selected by radio and multi-wavelength characteristics]
{\centering New Galactic Planetary nebulae selected by radio and multi-wavelength characteristics}

\author[V. Fragkou et al.]{V. Fragkou,$^{1,2}$\thanks{vfrag@hku.hk} Q. A. Parker,$^{1,2}$\thanks{quentinp@hku.hk}
I. S. Boji\v ci\'c,$^{3,1,2}$ N. Aksaker,$^{4,5}$
\\
$^{1}$Department of Physics, The University of Hong Kong, Hong Kong SAR, China\\
$^{2}$Laboratory for Space Research, The University of Hong Kong, Hong Kong SAR, China\\ 
$^{3}$Western Sydney University, Locked Bag 1797, Penrith South DC, NSW 1797, Australia\\ 
$^{4}$Vocational School of Technical Sciences, Cukurova University, Adana 01410, Turkey\\
$^{5}$Space Sciences and Solar Energy Research and Application center (UZAYMER), Cukurova University, Adana 01330, Turkey
\\}

\date{Accepted XXX. Received YYY; in original form ZZZ}

\pubyear{2018}

\begin{document}
\label{firstpage}
\pagerange{\pageref{firstpage}--\pageref{lastpage}}
\maketitle

\begin{abstract}

We have used the Cornish radio catalogue combined with the use of multi-wavelength data to identify 62 new Planetary Nebula (PN) candidates close to the Galactic mid-plane. Of this sample 11 have weak optical counterparts in deep narrow band H$\alpha$ imaging that allows their spectroscopic follow-up. We have observed eight of these candidates spectroscopically, leading to the confirmation of 7 out of 8 as PNe. All but one of our sample of newly detected PNe appear to be of Type~I chemistry with very  large [NII]/H$\alpha$ ratios.  This indicates that our selection method heavily favours detection of this kind of PN. Cornish is a low Galactic latitude survey where young objects and Type~I PNe (thought to derive from higher mass progenitors) are more plentiful, but where optical extinction is large. The very high success rate in correctly identifying PNe in this zone proves the efficacy of our radio and multiple multi-wavelength diagnostic tools used to successfully predict and then confirm their PN nature, at least in the cases where an optical counterpart is found and has been observed. The study reinforces the effective use of a combination of multi-wavelength and optical data in the identification of new Galactic PNe and especially those of Type~I chemistries whose dusty environments often prevents their easy detection in the optical regime alone.

\end{abstract}
\begin{keywords}
ISM: planetary nebulae-cornish-mir colour colour plots 
\end{keywords}



\section{Introduction - multi-wavelength characteristics of Planetary Nebulae}

Planetary Nebulae (PNe) research provides us with vital clues for understanding late stage stellar evolution of low-to-intermediate mass 
stars and the chemical enrichment of our Galaxy (see e.g. \citealt{Fr10}; \citealt{Par12a}). PNe are a short lived phase of late stage stellar evolution 
(a few tens of  thousands of years) and they emit most of their energy in narrow emission lines that makes them easily detectable in narrow band surveys and thus visible to greater distances in the Galaxy than their main-sequence counterparts. These strong emission lines also permit the 
determination of nebula abundances, expansion and radial velocities and, through photo-ionisation modeling, estimations of their central 
star (CS) temperatures. The study of the Galactic PN population across 1-9~Gyrs of stellar evolution can also trace Galactic star 
formation history (see \citealt{Buz06}).

The majority of PNe (more than 90\%; see e.g. \citealt{Fr10}) have been discovered through optical and narrow-band surveys 
(e.g. \citealt{Par05}; \citealt{Drew05}) and reported in \cite{Par06}, \cite{Mis08} and \cite{Sab14}. 
However, optical bands suffer from obscuration by interstellar dust along the line of sight, which may hide 
up to 90\% of PNe that reside on or close to the Galactic Plane and where extinction is significant (see e.g. \citealt{Jac04}; \citealt{Coh07a}; 
\citealt{Mis08}; \citealt{Jac10}; \citealt{Par12b}). Given this and the fact that the Galactic Plane is a highly crowded area, evolved 
low-surface brightness  PNe and distant and faint PNe are hard to detect at low Galactic latitudes 
(see e.g. \citealt{Sab10}). This limitation of the optical bands can be partially solved 
with observations at longer wavelengths like the Infrared (IR) and at radio frequencies, which are much less affected by dust extinction 
(e.g. \citealt{Zha12}; \citealt{Hoar12}). However, spectroscopic confirmation of PN candidates uncovered at longer wavelengths is usually 
problematic because many lack any obvious optical counterpart so spectroscopy at longer non-optical wavelengths would be needed. 

A multi-wavelength approach to PN study is also a key way to describe PN evolutionary diversity as different PN characteristics and mass 
components are better traced by different wavelengths (see \citealt{Kwok10}). Many past authors have used multi-wavelength 
measurements for the identification of PNe hidden in the optical and for their discrimination from other sources that may mimic PNe, like 
H\,{\sc ii} regions, Supernova Remnants (SNRs) and Symbiotic Stars (e.g. \citealt{Coh01}; \citealt{Coh03}; \citealt{Coh07a}; \citealt{Cor08}; 
\citealt{Coh11}; \citealt{Fr10}; \citealt{Par12b}; \citealt{And12}). Optical (H$\alpha$) emission traces the main PNe ionised gas component (see \citealt{Kwok08}). 
Radio continuum emission traces the closely related free-free emission from hot ionized gas that is present in both PNe and H\,{\sc ii} regions and is usually much 
stronger in the second case \citep{And12}. IR emission principally maps the nebular dust component though some MIR emission is from lines 
such as [OIV] at $24 \mu$m (the MIR analogues of high excitation lines in the optical 
such as He\,{\sc ii}).  At millimeter and sub-millimeter wavelengths emission can often be observed from molecules present in PNe  (see 
\citealt{Kwok00}; \citealt{Kwok08}) while H2 observations can also trace the molecular content in warm gas (e.g. \citealt{Gled17}). At high energies an 
X-ray continuum can be observed  in some PNe emerging from shocked inner bubbles (e.g. \citealt{Guer15}).

In Section~2 we briefly describe the Cornish radio survey that serves as the base data for this paper. In Section~3 we outline our selection 
criteria and in Section~4 the multi-wavelength diagnostic tools that have been applied to our candidates. Section~5 describes our spectroscopic 
observations required for the confirmation of the nature of our candidates and in section~6 we discuss our findings. Finally, in Section 7 we summarize our results.

\section{The Cornish radio survey}
The high-resolution, Very Large Array (VLA) Cornish radio survey (\citealt{Hoar12}; \citealt{Pur13}), mapped sources emitting at 5~GHz at Galactic 
longitudes l = 10$^{\circ}$ - 65$^{\circ}$ and latitudes $\mid$ b $\mid$ $\le$ 1$^{\circ}$ completing the Northern coverage of the GLIMPSE survey at radio 
wavelengths.  While 5~GHz radio observations are not sensitive to Hyper-compact H\,{\sc ii} regions \citep{Pur13} the Cornish survey is ideal in detecting 
compact and young PNe hidden by dust in the optical bands. Hoare et al. (2012) expected to detect around 1000 PNe with some of these being new 
detections currently either invisible or not currently known from optical data. At frequencies equal to or more than 5~GHz it will be safe to assume 
that most (but not all) PNe detected at these frequencies (including from the Cornish survey) will be 
optically thin at this frequency. Cornish detected PNe are expected to present 5~GHz flux densities from 5 to 50 mJy for PNe of angular sizes of 
a few arcseconds assuming typical distances of several kpc \citep{Hoar12}.  In general we expect these compact and likely mostly young 
PNe to be denser and have some optical depth.

In this work we used sources in the Cornish radio catalogue for identifying PNe candidates 
along the direction of the Galactic Plane that have been previously missed. Multiple multi-wavelength data selection criteria to uncover PN candidates 
(Fig. 1a and b) were used and various discriminatory tools were applied to demonstrate the power of such multi-wavelength diagnostics for uncovering 
new PNe by presenting confirmatory spectroscopic observations and measurements for 8 of the 11 identified PN candidates that have faint optical 
counterparts. Preliminary results were presented in \cite{Fragkou17}.

\begin{figure*}
\centering
\includegraphics[scale=0.97]{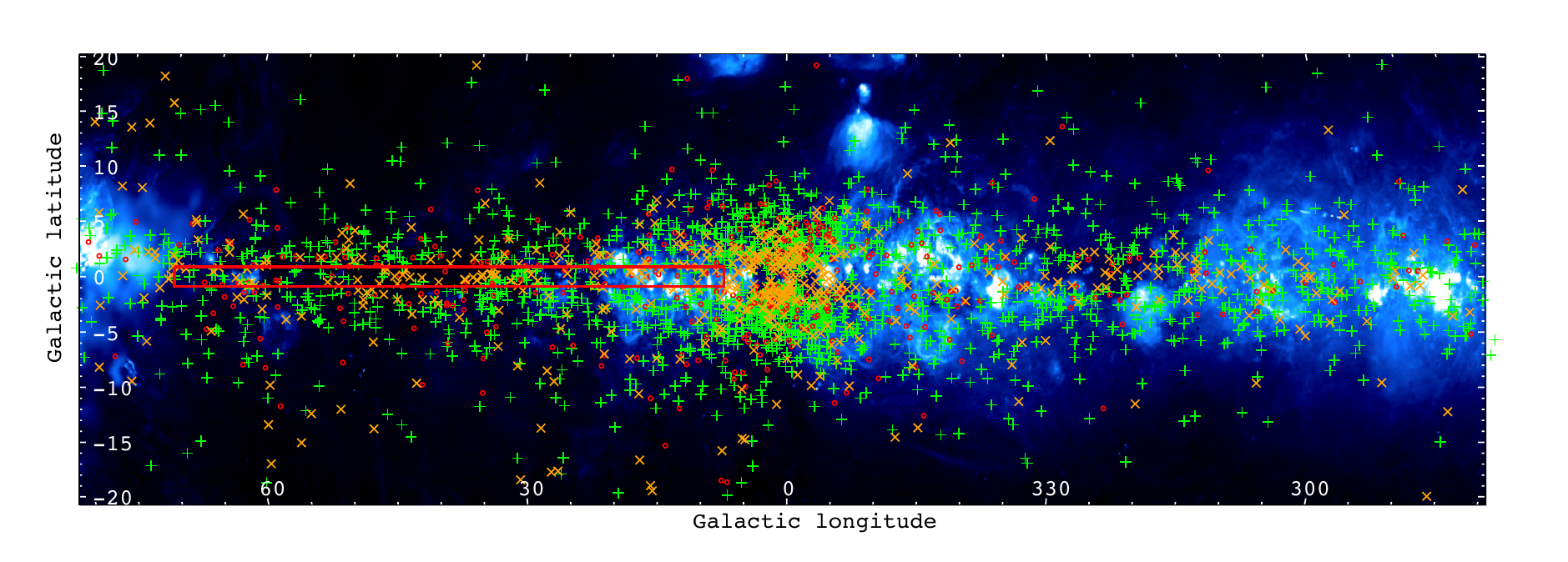}
\includegraphics[scale=0.95]{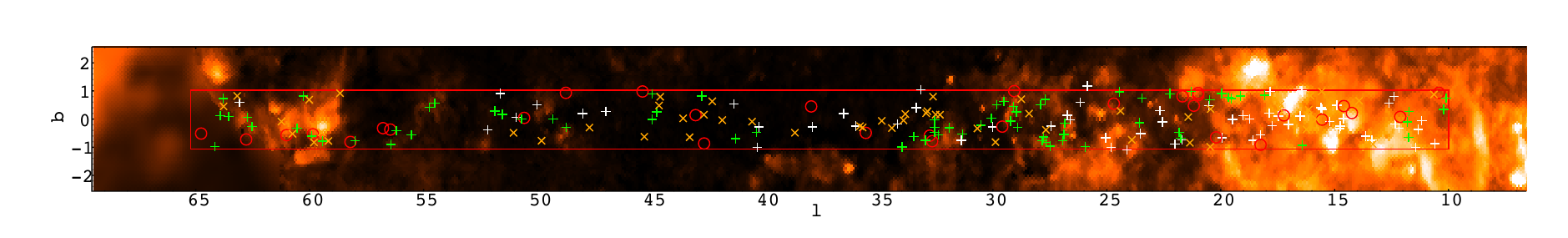}
\caption{Fig.1a (top): Inner Galactic Plane H$\alpha$ map superimposed with all known true (green), likely (orange) and possible (red) PNe identified as such from the HASH 
research platform (e.g. \citealt{Par16}). The inscribed small red rectangle indicates the area covered by the Cornish radio catalogue. Fig.1b (bottom) shows an expanded plot 
just of the Cornish zone with the same symbols as before. In addition, the PNe candidates uncovered by this work are indicated as white ``plus" signs.
The data for this diagram have been provided from Finkbeiner (2003) and Parker, Frew \&  Boji{\v c}i{\'c} (2016).}
\label{fig1}
\end{figure*}


\section{PN Candidate Selection}

For the selection of suitable PN candidates all 2637 objects from the Cornish radio catalogue (\citealt{Hoar12}; \citealt{Pur13}) as obtained from the CDS 
VizieR service  were cross-correlated with those that have IRAC counterparts (\cite{Pur13} quotes that 2638 sources are contained in the catalogue). 
No sample limits were placed on S/N (adopting the 7 sigma CORNISH catalogue limit \citep{Pur13}, peak flux or source angular diameter). Subsequently, 
the 1.4~GHz NVSS \citep{Con98} and 5~GHz Cornish radio fluxes {\bf ($S_{1.4GHz}$ and $S_{5GHz}$)} of all objects with NVSS detections available (again without imposing any S/N limits and 
adopting the 4 sigma NVSS catalogue limit \citep{Con98} were used for calculating their spectral indices $\alpha$ using the relation given in Anderson et al. (2011) below:

\smallskip
\[
\frac{S_{1.4 GHz}}{S_{5 GHz}}=\left(\frac{1.4 GHz}{5 GHz}\right)^{\alpha} \tag{1} \label{Eq. (1)}
\]
\medskip

The positional uncertainty allowed in the cross-correlation of Cornish with the NVSS data was 82~arcseconds and was based on the resolution of both 
surveys. The NVSS resolution is $\sim$80~arcseconds and that of Cornish $\sim$1.5~arcseconds.  The mean separation of matched sources between Cornish and NVSS 
was 10.6~arcseconds (median= 3.7~arcseconds). As stated, we expect most PNe detected in a 5~GHz survey, like Cornish, to be 
optically thin at that frequency with a spectral index around -0.1 expected between $S_{1.4GHz}$ and $S_{5GHz}$ (\citealt{Hoar12}; \citealt{Pur13}). Hence, all candidates with no NVSS data or with a spectral 
index $\alpha < -0.5$ were excluded from our selection. This excludes non-thermal radio sources such as Supernova remnants. 
The flux uncertainty was not accounted for in detail in the spectral index calculation but the limit for selected sources was $\alpha < -0.5$. 
For optically thin sources we expect a value around -0.1, this conservative approach takes into account the 
errors of the measured fluxes and the likely presence of sources that are also more optically thick. 
The maximum estimated error of $\alpha$ of all cross-correlated sources with $\alpha < -0.5$ is around 0.1 ($\sim20\%$).

PNe are expected to have GLIMPSE mid-infrared [8.0~$\mu$m] - [24~$\mu$m] and [5.8~$\mu$m] - [24~$\mu$m] colour indices ranging from 3.4 to 
8.7 and from 5.4 to 10.3 mag respectively \citep{Phil11}. A cross-correlation of our candidates with the longer mid-infrared wavelength and lower resolution 
MIPSGAL source data at 24$\mu$m and 70$\mu$m \cite{Gut15}, and again without putting any S/N limits, allowed the exclusion of objects with MIPSGAL detections whose 
MIR colour indices lie outside these GLIMPSE ranges. The positional uncertainty allowed in the cross-correlation of the Cornish sources with MIPSGAL sources was set to a generous 
40~arcseconds, more than accounting for the resolution of both the MIPSGAL (6~arcseconds at 24$\mu$m) and Cornish surveys.  Only 16 possible matches were found within 
these lax allowed positional uncertainties and none when more conservative limits were imposed. As we show later, 7 out of 8 of the 11 candidates with optical counterparts 
(from all those uncovered via our original Cornish radio selection and then combined multi-wavelength selection criteria) and observed spectroscopically are confirmed PN. 
This suggests source confusion, if present, is not significant.

Following \cite{Hoar12}, we excluded all objects with 5~GHz 
radio fluxes more than 110~mJy. They expected PNe at typical distances found via the Cornish survey to have 5~GHz radio flux densities of less than 50~mJy as 
much larger fluxes indicate H\,{\sc ii} regions \citep{Fil09}.  Indeed, within our final radio selection below, only 4 have 5~GHz fluxes in excess of 50~mJy. 
It is still possible to miss a few PNe even with this conservative limit but an even larger one would add many more mainly 
H\,{\sc ii} regions to our sample. This would require significant amounts of additional spectroscopic follow-up with very little likely return. The very high success rate of this 
adopted position (the median 5~GHz radio flux of our confirmed PNe is 9.7~mJy with a maximum of 33.9~mJy while the median of the full sample of 62 objects is 16.09~mJy) 
indicates that an upper flux limit a bit larger than double the flux expected from \cite{Hoar12} is a very good basis for primary selection of the best candidates. 

In detail, 1463 Cornish objects are found to have NVSS counterparts and from these around 150 were excluded on the 
basis of their large 5~GHz fluxes ($>$110~mJy).  A further 719 objects were excluded as a result of the imposed spectral index limit leaving 594 candidates. The 
cross-correlation of these remaining objects with the GLIMPSE catalogue resulted in 322 candidates.  From these 11 were excluded on the basis of their [8.0~$\mu$m] - [24~$\mu$m] 
and [5.8~$\mu$m] - [24~$\mu$m] colour indices being clearly outside those found for known PNe. 
Finally, the resulting 311 radio-selected PN candidates were checked individually for extant literature data that might indicate clues as to their nature. Already 
identified objects (22) that most likely are not misidentifications were removed from the list (i.e. objects like H\,{\sc ii} regions and emission line stars were not removed as these 
could possibly be misidentifications). Of those removed 12 were known PNe, while the remaining 10 were identified as a pulsar, 3 masers, a galaxy 
and 5 stars. As this is a Galactic Plane survey stellar source confusion would always be an issue so that false stellar matches within the 5~arcsecond matching criteria 
are always possible though only the brighter stars in any 5~arcsecond region would have a current SIMBAD entry.

The available multi-wavelength colour composite images of the remaining 289 candidates were visually examined one by one using the HASH 
research platform \citep{Par16} where up to 40 different multi-wavelength images and combinations are available from the UV to radio regimes for each source. Many are 
produced as ``RGB" false colour images of various optical, IR and MIR combinations (see Parker et al. 2006 and 2017 for details). Particular attention 
was paid to matches with the narrow-band H$\alpha$ optical data and the high resolution UKIDSS NIR data \citep{Law07}. These RGB multi-wavelength image combinations are 
powerful diagnostic tools in their own right as described in  \cite{Par12b}  - refer section 4 of that paper.

Unfortunately, the majority of our PN candidates are not detected in the available optical 
and narrow-band images. However, based on the other non-optical, multi-wavelength diagnostic indicators we have established (e.g. \citealt{Fr10}, \citealt{Par12b} and 
\citealt{Coh11}) we assessed that 62 of the 289 radio selected candidates have a high chance to be PNe. This includes the example shown in Fig. 2 ``FBP2",  the 
first spectroscopically confirmed PNe from our sample listed in Table~1. We adopt the PN naming convention that has been employed by \cite{Par06} for the MASH PNe survey 
and \cite{Par16} for the HASH PNe surveys of the initials of the surnames of the authors and discoverers, in this case Fragkou, Bojicic and Parker. 
The radio data of all candidates are presented in Table~1 which includes accurate equatorial radio positions (J2000 RA/DEC), 5~GHz flux determinations with errors
and the estimated radio angular diameters. All but 6 of these sources have angular sizes less than 3~arcseconds. From the table it is clear that a significant fraction of objects
have $S_{1.4GHz} < S_{5GHz}$ and so seem to have some optical depth at 1.4GHz.
We find that 11 of these 62 tabulated colour selected candidates have weak optical counterparts with seven in the SHS \citealt{Par05} Southern Galactic Plane and just 4 solely 
in the IPHAS \cite{Drew05} Northern Galactic plane surveys. These optical detections allows their direct spectroscopic follow up. The 8 sources subsequently observed 
spectroscopically to date are placed at the head of Table~1 and refer to, in order of increasing RA, as FBP2, FBP3, FBP4, FBP6, FBP7, FBP8, FBP9 and FBP11 and explicitly 
tabulated again in the spectroscopic observation log in Table~3. The remaining 3 sources with optical counterparts are FBP1, FBP5 and FBP10 also listed at the head of Table~1.    

Apart from an initial examination of the 2MASS near infrared (NIR) counterparts the UKIDSS NIR survey was also examined as these newer higher resolution NIR images 
enabled just resolved (not stellar) sources to be identified. Our previous published work, e.g. \cite{Par12b}, has shown that PNe tend to have distinctive combined 2MASS JHK colours and this 
is also true for UKIDSS which also uses the same JHK filters. Effectively every UKIDSS source had a 
2MASS  counterpart but not all 2MASS sources had a UKIDSS equivalent due to some gaps in coverage. Of the 289 candidates only 8 UKIDSS additional sources were identified 
as possible  PNe and they are listed  separately in Table~2 along with their radio parameters. Only 1 candidate has a 5~Ghz radio flux $>$50~mJy. None of these additional 8 
candidates had an optical counterpart. As these objects require further  examination and more available data they have not been included in the following analysis and are not in the 
sample of 62 referred to throughout this paper.

We did not cross-correlate the radio selected sample with optical catalogues as we are 
looking for very faint emission candidates in narrow-band optical surveys and these are currently mostly uncatalogued. As the final candidate sample was modest in size 
and of small angular extent we simply looked at the astrometric Cornish and narrow-band survey images for very close source positional co-incidences. The 
positional match in each case was very clear (within $\sim$1~arcsecond). These optical surveys have excellent astrometric integrity and positional uncertainties 
are typically $<$0.5~arcseconds  (refer to the relevant publications). The SHS has 0.3~arcseconds astrometric error, \cite{Par05} and IPHAS 0.5~arcseconds, 
\cite{Drew05} with 100~mas proper-motion accuracy (Gonzalez-Solares 2008). All selected narrow-band optical source matches are also very compact (only
$\sim$2~arseconds across). A significant difference between the radio and optical dimensions for PNe is not expected. 
If these sources are mostly PNe this indicates that they are both distant and young. 

\begin{table*}  
\centering
\caption[]{J2000 equatorial positions (to $\sim$0.5~arcseconds), Cornish and NVSS radio fluxes, estimated flux errors, estimated 
spectral index values (with errors) and radio determined angular diameters for the 62 PN candidates revealed 
from this multi-wavelength study. The 11 objects (including 7 confirmed PNe) with optical counterparts are presented in first part of the table. 
Radio diameters are deconvolved FWHM values from the Cornish catalogue.}  
\label{table1}
\begin{tabular}{cccccccccc}  
\noalign{\smallskip}  

ID & RA J2000 &Dec J2000 & 5~GHz flux & 5 GHz flux error & 1.4 GHz flux & 1.4 GHz flux error &spectral & $\alpha$  & radio diam.\\
 & (h:m:s) & (\degr\:\arcmin\:\arcsec)  & (mJy) &  & (mJy) & & index $\alpha$ & error & (\arcsec) \\
 \hline
FBP1&18:10:25.11&-19:18:00&102.17&12.73&25.8&1.4&1.08&0.46&10.60 \\
FBP2&18:13:12.13&-18:20:24&33.65&3.3&-&-&-&-&1.80 \\
FBP3&18:15:53.22&-15:45:36&8.84&1.22&-&-&-&-&1.62\\
FBP4&18:19:47.16&-12:57:00&15.37&1.76&2.8&0.6&1.34&1.40&1.90 \\
FBP5&18:25:10.01&-12:18:36&6.41&1.52&-&-&-&-&2.30 \\
FBP6&18:40:06.71&-7:42:00&9.69&1.39&-&-&-&-&1.40 \\
FBP7&18:42:15.85&-5:00:36&5.47&0.99&-&-&-&-&1.75 \\
FBP8&19:04:57.33&07:44:24&33.9&3.45&13.3&0.6&0.74&0.16&2.30 \\
FBP9&19:21:27.76&15:22:48&19.51&1.85&4.3&0.5&1.19&0.64&1.75 \\
FBP10&19:23:07.17&16:59:24&18.85&1.76&4.3&0.5&1.16&0.60&1.65 \\
FBP11&19:48:23.27&26:49:12&5.23&1.02&-&-&-&-&1.50 \\
\hline
FBP12&18:09:41.62&-17:49:12&19.59&1.98&3.9&0.5&1.27&0.82&1.70 \\
FBP13&18:11:01.01&-17:44:24&17.11&1.7&19.9&5&-.12&0.02&1.57 \\
FBP14&18:11:52.32&-19:18:36&13.82&2.41&-&-&-&-&2.30 \\
FBP15&18:13:23.74&-18:33:36&11.97&1.42&-&-&-&-&1.10 \\
FBP16&18:14:27.14&-19:31:12&14.58&2&7.6&0.7&0.51&0.13&1.70 \\
FBP17&18:16:59.73&-14:10:12&16.41&2.68&7.8&0.6&0.58&0.17&2.60 \\
FBP18&18:16:59.81&-16:15:00&4.39&1.09&-&-&-&-&1.10 \\
FBP19&18:17:21.09&-17:23:24&19.91&3.28&-&-&-&-&3.50 \\
FBP20&18:17:23.12&-17:42:36&9.68&1.29&4.3&1&0.64&0.30&1.70 \\
FBP21&18:17:33.04&-15:12:00&21.91&2.13&8.5&0.7&0.74&0.19&1.66 \\
FBP22&18:17:41.95&-15:16:12&21.92&3.22&-&-&-&-&1.40 \\
FBP23&18:18:17.07&-16:14:24&19.77&2.57&-&-&-&-&2.50 \\
FBP24&18:18:34.31&-15:46:12&10&1.62&-&-&-&-&1.50 \\
FBP25&18:18:38.07&-16:15:36&18.25&2.47&-&-&-&-&1.80 \\
FBP26&18:20:22.87&-14:31:48&15.78&2.3&-&-&-&-&3.50 \\
FBP27&18:22:13.53&-13:41:24&9.42&1.74&-&-&-&-&2.50 \\
FBP28&18:22:29.82&-14:13:12&13.32&2.31&-&-&-&-&1.90 \\
FBP29&18:22:40.49&-13:16:48&23.35&2.16&-&-&-&-&1.57 \\
FBP30&18:24:03.46&-13:36:36&59.11&5.99&43.4&1.4&0.24&0.03&4.70 \\
FBP31&18:25:04.14&-12:37:48&107.46&10.62&-&-&-&-&1.40 \\
FBP32&18:25:14.95&-20:11:24&34.54&3.84&-&-&-&-&1.70 \\
FBP33&18:26:11.16&-13:18:00&8.24&0.91&-&-&-&-&1.50 \\
FBP34&18:26:34.33&-11:58:12&18.81&1.85&-&-&-&-&1.10 \\
FBP35&18:26:47.26&-10:48:36&15.62&1.88&10.5&0.7&0.31&0.05&1.90 \\
FBP36&18:27:31.92&-13:05:60&7.6&1.5&9.7&2.6&-.19&0.04&2.40 \\
FBP37&18:29:47.85&-11:51:36&14.12&1.38&-&-&-&-&1.50 \\ 
FBP38&18:31:28.54&-9:00:00&14.15&1.42&3.5&0.6&1.10&0.69&1.68 \\
FBP39&18:32:42.81&-9:16:48&32.09&3.03&-&-&-&-&1.30 \\
FBP40&18:33:22.39&-10:20:24&36.49&6.02&-&-&-&-&4.70 \\
FBP41&18:34:16.58&-5:45:36&31.34&4.35&22.3&0.8&0.27&0.04&2.80 \\
FBP42&18:34:29.01&-10:07:48&14.22&1.65&11.9&2&0.14&0.03&1.40 \\
FBP43&18:36:00.12&-8:37:12&27.22&2.69&8.8&0.7&0.89&0.27&1.70 \\
FBP44&18:36:55.64&-5:45:00&12.38&2.18&7.1&0.6&0.44&0.12&2.60 \\
FBP45&18:39:09.73&-8:21:36&16.4&1.55&10&0.6&0.39&0.06&1.50 \\
FBP46&18:39:21.51&-7:19:12&30.11&2.75&5.3&0.8&1.36&1.07&1.74 \\
FBP47&18:39:35.49&-5:28:12&13.16&1.51&-&-&-&-&1.20 \\
FBP48&18:39:55.87&-5:39:00&69.92&7.45&70.6&2.7&-.01&-&2.20 \\ 
FBP49&18:47:04.71&-2:43:48&4.54&0.94&-&-&-&-&1.62 \\
FBP50&18:48:06.96&00:40:12&25.72&2.36&32&1.1&-.17&0.01&1.50 \\
FBP51&18:50:45.53&00:33:36&2.62&0.65&-&-&-&-&1.50 \\
FBP52&18:51:16.12&-1:43:48&22.66&2.29&-&-&-&-&1.60 \\
FBP53&18:54:20.74&01:01:48&10.14&1.8&3.7&0.5&0.79&0.38&2.30 \\
FBP54&18:57:18.51&03:18:00&13.4&2.55&13.6&0.7&-.01&-&3.20 \\
FBP55&18:57:58.33&02:37:12&18.18&1.7&-&-&-&-&1.66 \\
FBP56&19:01:30.32&04:18:00&32.54&2.99&-&-&-&-&1.64 \\
FBP57&19:05:50.88&06:23:24&7.03&0.99&-&-&-&-&1.50 \\
FBP58&19:08:36.75&06:07:12&9.12&1.48&9&0.6&0.01&-&1.90 \\
FBP59&19:16:29.08&12:36:00&18.06&1.91&4.9&0.5&1.02&0.44&1.10 \\
FBP60&19:18:40.35&13:27:36&6.03&0.91&5&0.5&0.15&0.03&1.65 \\
FBP61&19:23:07.17&16:59:24&18.85&1.76&4.3&0.5&1.16&0.60&1.65 \\
FBP62&19:28:56.07&16:51:36&22.89&2.08&36.7&1.2&-.37&0.02&1.55 \\

\end{tabular}
\end{table*}  

\begin{table*}  
\centering
\caption[]{Cornish J2000 equatorial positions (the median uncertainty in RA and Dec is 0.12 \degr ), 
Cornish and NVSS radio fluxes and their errors, spectral index estimates 
and radio determined angular diameters for the 8 sources of which only their UKIDSS counterpart images imply a possible PN. 
This includes several that appear resolved in UKIDSS (indicated with an asterisk).}
\label{table2}

\begin{tabular}{cccccccccc}  
\noalign{\smallskip}  

\hline

ID & RA J2000 &Dec J2000 & 5~GHz flux & 5 GHz flux & 1.4 GHz flux & 1.4 GHz flux &spectral & $\alpha$ & radio diam.\\
 & (h:m:s) & (\degr\:\arcmin\:\arcsec)  & (mJy) & error & (mJy) & error & index $\alpha$ & error & (\arcsec) \\
\hline
FBP63 &18:39:09.70&-8:21:49&16.4&1.55&10.2&0.6&0.37&0.05&1.50* \\
FBP64&18:46:17.30&-1:40:22&49.48&4.55&5.8&0.6&1.68&1.56&2.21*\\
FBP65&19:02:05.60&05:40:31&62.27&6.41&11.5&0.6&1.33&0.65&1.95 \\
FBP66&19:13:36.00&10:21:14&14.02&1.48&8.8&0.5&0.37&0.05&1.59 \\
FBP67&19:46:10.50&26:28:01&14.27&1.36&10&0.5&0.28&0.03&1.50* \\
FBP68&19:47:07.60&26:28:25&5.37&0.69&4.5&0.5&0.14&0.02&1.50 \\
FBP69&19:49:11.80&26:10:52&6.97&0.8&4.1&0.5&0.42&0.09&1.56 \\
FBP70&19:58:38.20&27:38:24&5.57&0.66&4.8&0.6&0.12&0.02&1.50 \\
\hline

\end{tabular}
\end{table*}

Based on previous work reported in the literature by our group, e.g. see \cite{Drew05}; \cite{Coh07a}; \cite{Coh11} and  \cite{Par12b}, 
we have found that assigning R (red), G (green) and B (blue)  image channels can be very effective diagnostically
in revealing promising PN candidates as shown in the following multi-wavelength colour images of one of our objects FBP2 (Fig.2). The channels are combined naturally with unit 
weighting and no enhancement of one channel over the other is made.

The eye (with apologies to those who are colour blind) is an excellent discriminator in this sometimes subtle process compared to solely relying on 
numerical photometric colour ranges. Furthermore, visual inspection can also reveal if a source is resolved (and so is a nebula and not a star) and is not a deblended element of a 
diffraction spike around a star (see examples in Parker et al. 2012b).

Verified PNe of moderate to low surface brightness present as violet 
or pink-red colours in the NIR 2MASS and higher resolution UKIDSS J, H and Ks  band images, \cite{Skr06}, \cite{Law07} due to the combinations of NIR emission lines that PNe 
produce at these wavelengths such as the recombination lines of Hydrogen Bracket series and He I. They present as red sources in the standard WISE321 12, 4.6 and 3.4~$\mu$m 
MIR band images - \cite{Wr10}; yellow in WISE432 22, 12 and 4.6~$\mu$m MIR band images, \cite{Wr10} again to due emission lines such as [OIV] and red, orange or violet in 
IRAC432 8.0, 5.8 and 4.5~$\mu$m band images, see \cite{Benj03}; \cite{Chur09}. IRAC bands 3 and especially 4  sample the PAH emission, often seen in longer wavelength PNe 
spectra (Cohen et al. 2007a) showing an association with carbon-PNe (Cohen et al. 1989), while the IRAC 2 band is sensitive to their H2 molecular line emission (Cohen et al. 2007b) 
Finally, they present as red in SHS H$\alpha$, short-red and broadband blue filter composite optical images due to the dominance of H$\alpha$ that is assigned to the R channel, 
\cite{Par05}; \cite{Par12b} and likewise for the IPHAS H$\alpha$ and broadband r' and i', \cite{Drew05} false-colour images. 

\begin{figure*}
\centering
\includegraphics[scale=0.4]{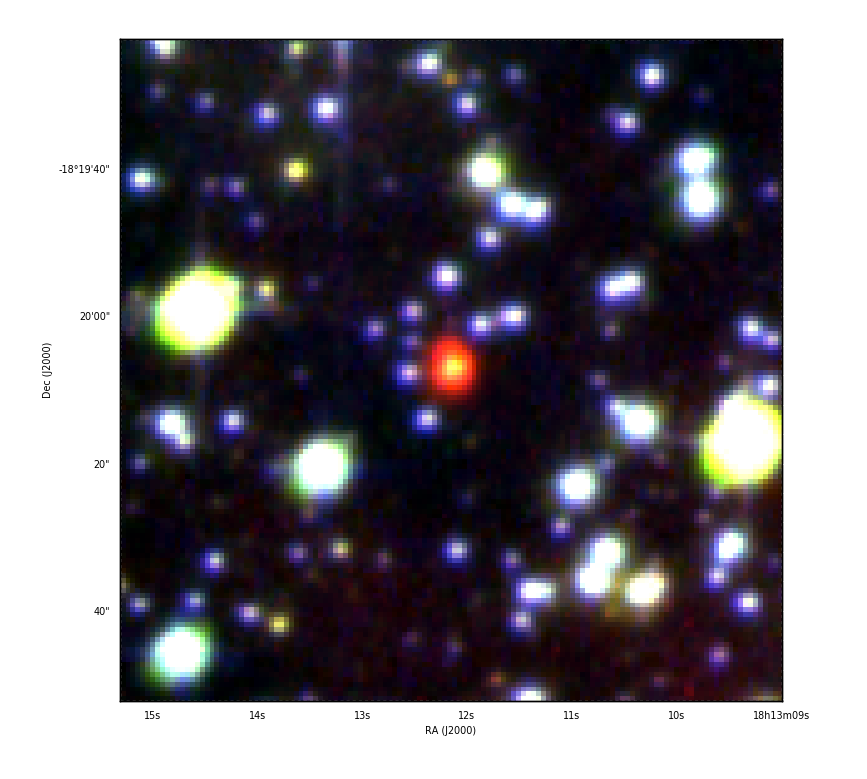}
\includegraphics[scale=0.4]{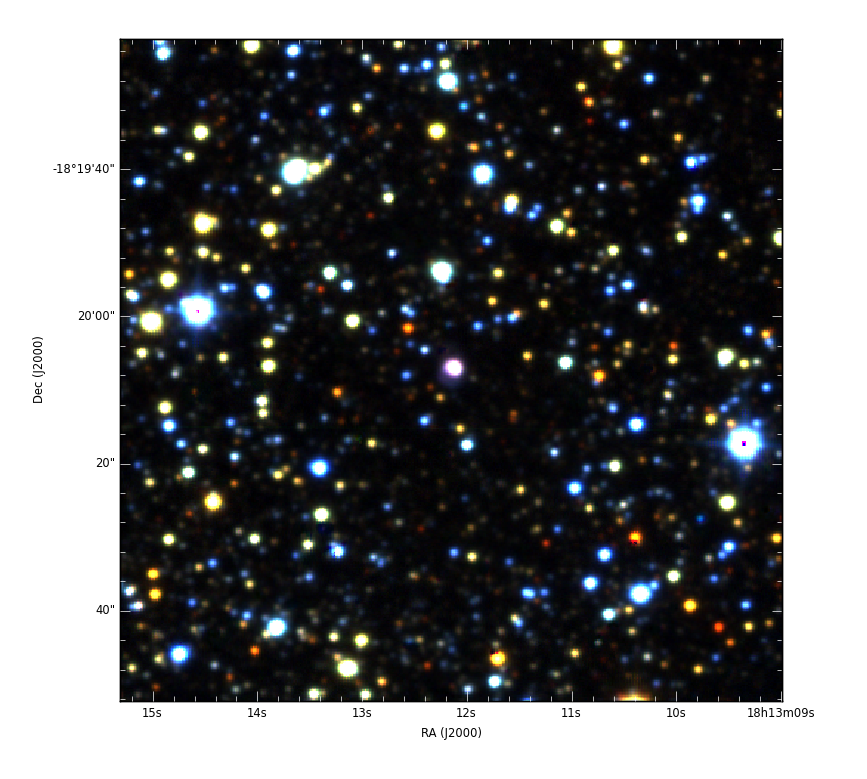}
\includegraphics[scale=0.4]{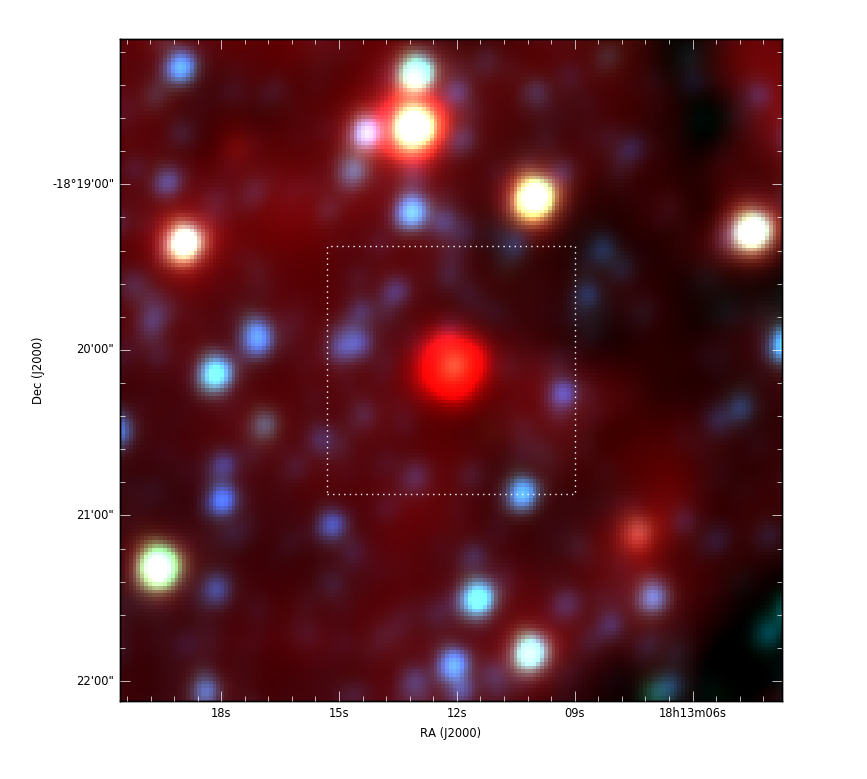}
\includegraphics[scale=0.4]{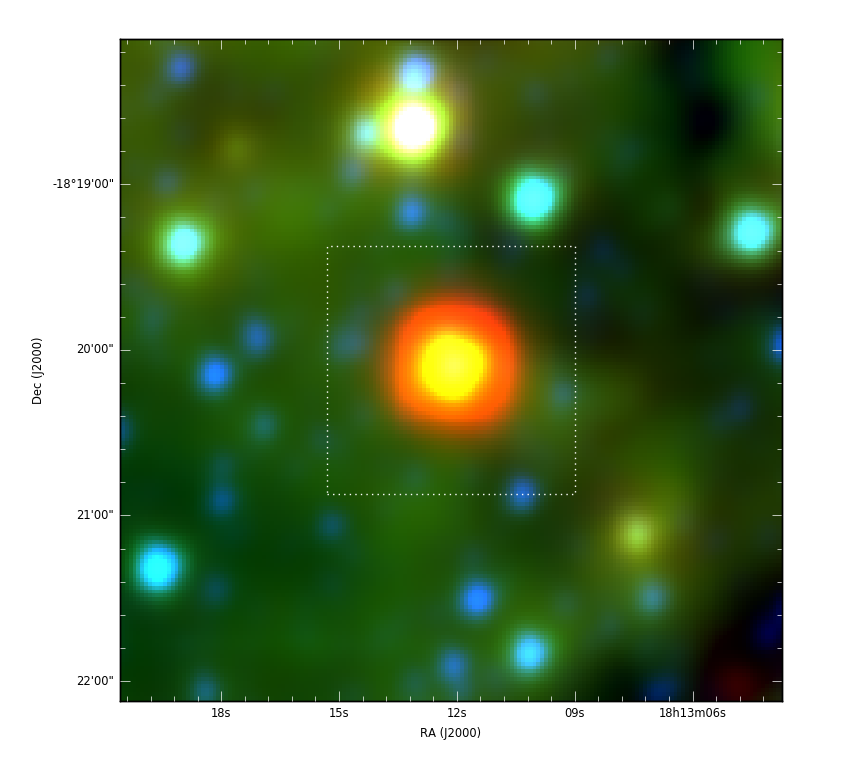}
\caption{SHS H$\alpha$, R and Bj band (top left), UKIDSS JHK (top right), WISE321 (bottom left) and WISE432 (bottom right) 
colour-composite ``RGB" images of the first promising PN radio selected candidate selected in this work (designated FBP2) from the visual inspection 
of its multi-wavelength images. It has now been confirmed spectroscopically (see section 4) as a PN. In each combination the R, G and B channels 
are assigned in the same order as listed above. It is the first object listed in Table~1. Individual survey images were provided from the HASH research platform \citep{Par16}.}
\label{fig2}
\end{figure*}

For further clarifying the nature of our 62 candidates and their likelihood to be PNe we also employed multi-wavelength photometric diagnostic tools that have 
been successfully used in the past for PNe identification (see e.g. \citealt{Coh11}; \citealt{Par12b}). The 8.0$\mu$m/848~MHz fluxes of our candidates were over-
plotted (with big star symbols whose colours indicate their MIR/radio flux ratios) on an IRAC [5.8]-[8.0] versus [3.8]-[4.5] colour-colour plot (Fig. 3). Following 
Anderson et al. (2011) the MIR/radio flux ratios of our candidates were calculated after converting our Cornish 5~GHz to 848~MHz radio fluxes using the relation: 

\smallskip
\[
\frac{S_{848 MHz}}{S_{5 GHz}}=\left(\frac{0.848 GHz}{5 GHz}\right)^{-0.1} \tag{2} \label{Eq. (2)}
\]

\medskip

\noindent for optically thin sources (see \citealt{And11}; \citealt{Hoar12}), a condition we adopted from our selection criteria. 

The MIR/radio ratio is a useful discriminator between PNe and HII regions (see Cohen \& Green 2001).
Our objects (indicated by red filled squares) were also plotted on 2MASS H-Ks versus J-H colour-colour plots 
(Fig. 4). Representatives of the most common PNe mimics (e.g. from \citealt{Acker87}; \citealt{Par06}; \citealt{Fr10}), with coordinates obtained from published 
catalogues (\citealt{Kur94}; \citealt{Giv05}; \citealt{And14}; \citealt{Par16}), are also plotted on all of our colour-colour plots for illustrative purposes. We selected the 
same types of mimics as \cite{Coh11}.  Nine of our 62 candidates that have no measurement for all individual IRAC bands  cannot be included in the corresponding IRAC 
colour-colour plot. Measurements of all 2MASS bands exist in published data for only 23 of our  62 candidates (refer 2MASS All-Sky Catalog of Point Sources, 
\citealt{Cut03}), but most of these have poor quality flags in at least one of the filters. We only include those 5 candidates with high quality flags in our 
2MASS colour-colour plot. We note several 2MASS sources have no UKIDSS counterpart due to missing coverage. 

The IRAC colour-colour plots in Fig. 3 shows our candidates  generally fall 
in or around the black `P' boxes, which indicate the areas centered on the median colour-colour values of confirmed PN taken from \cite{Coh11}. The box extent is 
given by 3 standard errors on the mean of the colours of such confirmed PNe. Sources located in and around this box  
are distinctive from most other identified object types. The plot where the tightest newly confirmed PN grouping appears is Fig.3, the IRAC [5.8]-[8.0] versus [3.8]-
[4.5]  colour-colour plot, but interestingly they are not inside the box. 
This indicates that in our flux and spectral index limited radio samples from matched Cornish and NVSS sources the IRAC colour-colour selection should be 
extended in the  [5.8]-[8.0] band selection from a range of approximately 1.3 to 2.4 and in the [3.8]-[4.5]  band selection from 0.6 to 1.8. 
These new colour-colour ranges encapsulate $\approx$ 90\% of the radio selected sample and all of the confirmed new PNe. 
As we shall see below, objects in this zone may sample a different PN population type.

Based on published data (see \citealt{Coh11}) diffuse H\,{\sc ii} regions fall in 
the blue dH boxes (also set by the median colour value $\pm$ 3 sem). However, such diffuse H\,{\sc ii} regions are scarcely found within our radio selected 
samples (\citealt{Kur94}; \citealt{Giv05}; \citealt{And14}), which mainly consists of compact and ultracompact H\,{\sc ii} regions that may be the most serious 
contaminant among our candidate objects (see \citealt{Coh11}; \citealt{And12}). It is also evident that more than half of our candidates have MIR/radio flux ratios 
between 0.5 and 10, which according to Cohen et al. (2007b, 2011), further supports their PNe nature. The 2MASS colour-colour plot in Fig.4. shows most of our 
candidates with 2MASS data (only 4 out of 62 sources)  are distributed along the horizontal (H-Ks) axis, distinguished from objects of different nature and being in 
generally good agreement with the findings of \cite{Cor08} for 2MASS colour indices of PNe. 

Determining any morphological detail for all candidates with optical counterparts is difficult as they are all compact emitters with none being more than a few arcseconds in 
diameter, indicating either youth and/or distance. In the Cornish images most of our candidates are very compact, while in the WISE images they mostly present 
a round shape as they are unresolved in these MIR surveys. Some of the UKIDSS images are resolved but their shapes are unclear beyond recognising they 
are nebulae and perhaps in some case elliptical.

\begin{figure*}
\centering
\includegraphics[scale=0.5]{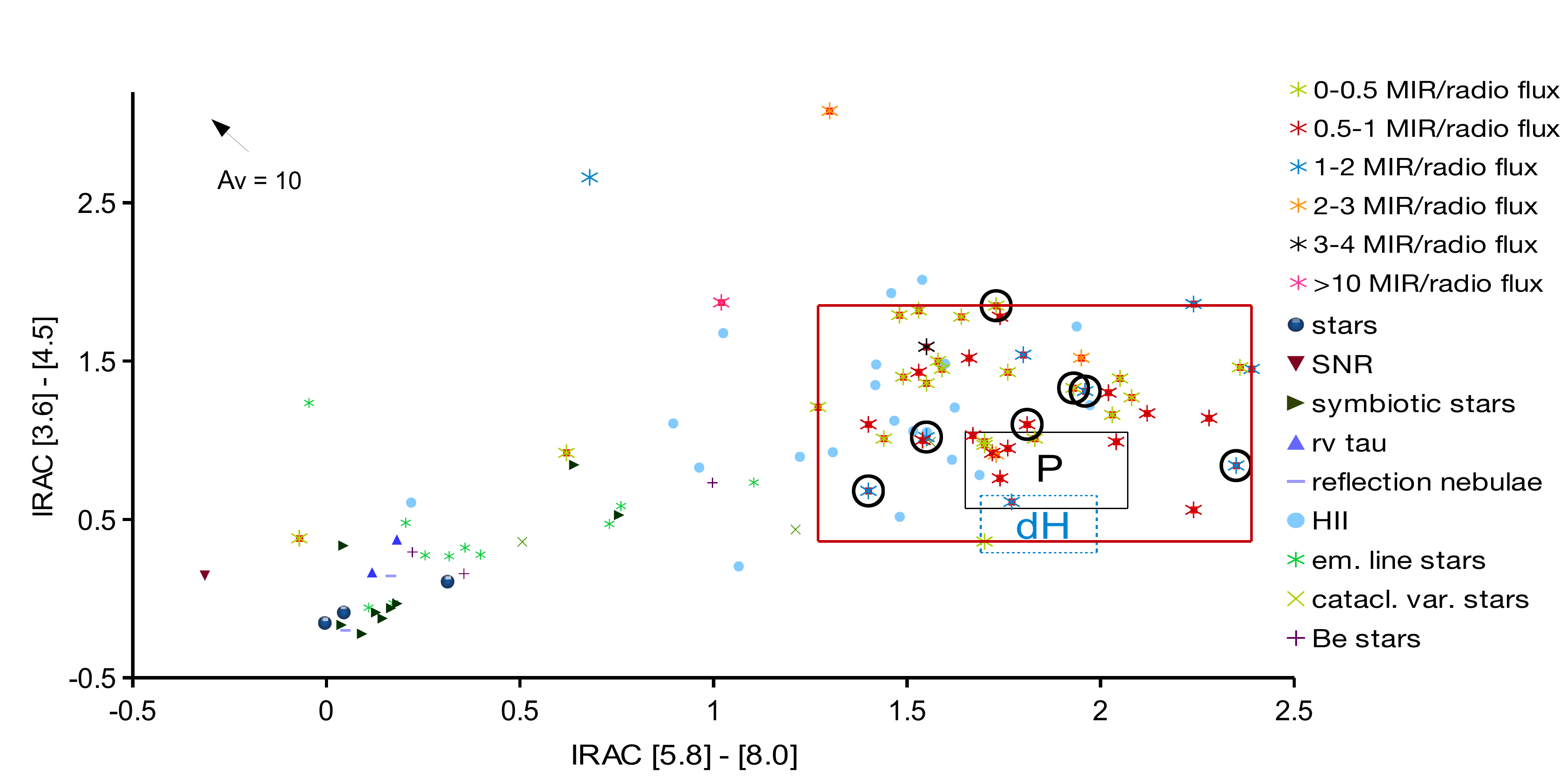}
\caption{MIR (8.0 $\mu$m)/radio (843MHz) fluxes of our PNe candidates over-plotted on an IRAC [5.8]-[8.0] versus [3.6]-[4.5] colour-colour plot. Green, red, blue, 
orange, black and pink large star symbols show our PN candidates with MIR/radio flux ratios ranging from 0 to 0.5, 0.5 to 1, 1 to 2, 2 to 3, 3 to 4 and larger than 10 
respectively. The black 'P' and the blue 'dH' boxes indicate the areas where most previously detected PNe and diffuse H\,{\sc ii} regions are located (see \citealt{Coh11}), 
while symbols for objects of different nature are indicated in the legend. A new red box indicates the suggested region in this colour-colour plane for PNe. The open black circles 
indicate our confirmed PNe. Coordinates of object types other than our PNe candidates are obtained from published catalogues (\citealt{Kur94};\citealt{Giv05}; \citealt{And14}; \citealt{Par16}) 
and cross-correlated with the GLIMPSE catalogue \citep{Spitz09} for finding their IRAC colours.}
\label{fig3}
\end{figure*}

\begin{figure*}
\centering
\includegraphics[width=\textwidth]{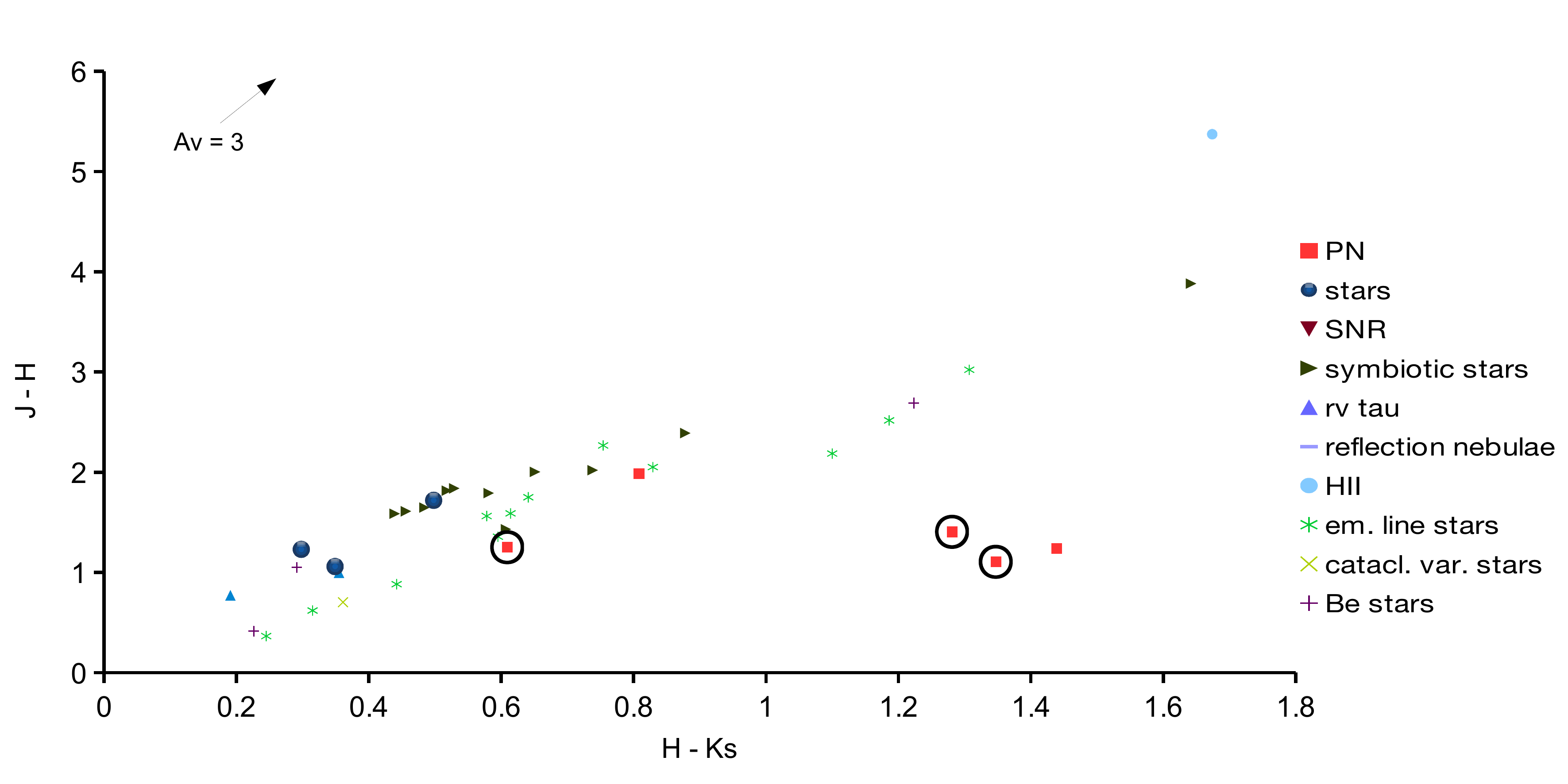}
\caption{Our PN candidates among frequent PN mimics plotted on a 2MASS H-Ks versus J-H colour-colour plot. References and symbols are as in 
Figures 3. All confirmed PNe here fall in the lower bound regions occupied by previously confirmed PNe.}
\label{fig4}
\end{figure*}

\section{Spectroscopic Confirmation of PNe Candidates}

Eight of the 11 PNe candidates that have optical counterparts have now been observed spectroscopically on 1.5 to 2-m class telescopes. The three remaining are 
the most optically faint PNe candidates that require observations with a larger telescope. A brief log of our observations is presented in Table~3 and they are 
described in detail in the following subsections.

\subsection{Observations}

Low-dispersion, long-slit spectra were obtained for eight of our candidates. Seven were observed with the newly commissioned SpUpNIC spectrograph (see \citealt{Cra16}) attached 
to the 1.9 m Grubb Parson telescope at South African Astronomical Observatory (SAAO) on May 24 and 27, 2017 and one with the TFOSC spectrograph on the 1.5 m 
Ritchey-Cretien telescope at TUBITAK National Observatory of Turkey (TUG) on July 04, 2017. For the SAAO run, the $2048\times512$ (13.5$\mu$m pixel) E2V 
CCD42-10 CCD and low dispersion grating were used covering $\sim$4000 to 9550\AA. For the TUG run we used the $2048\times2048$ (15$\mu$m pixel side) 
Fairchild 447BI CCD along with a low-dispersion grism covering the wavelength region from 3230 to 9120\AA. The extra coverage and system sensitivity out 
towards one micron with the new SpUpNIC spectrograph grating combination allows the strong far red [SIII] PNe nebular lines to now be observed. The exposure 
times ranged from 900 to 3600 seconds (see Table 2) depending on the faintness of the candidates. In the case of the grism used with the TFOSC spectrograph 
the resolution was R=749, while for the G7 grating used with SpUpNIC spectrograph the resolution was R=700. The coordinates of the centers of the slit for each of 
our observations are given in Table 3. The slit width of SpUpNIC was adjusted at 2.1~arcseconds and of TFOSC at 2.4~arcseconds, while their slit lengths and 
orientations were 2~arcminutes East-West and 6.3~arcminutes North-South respectively. For reducing our data we adopted standard IRAF techniques. For flux 
calibration we observed the spectrophotometric standard stars LTT 4816, LTT 4364 and HR 8634 \citep{Ham92}. For the extraction of the sky background, areas 
free of field stars were chosen around the centre of the slit as close as possible to the compact nebulae spectra. 

\begin{table*}  
\caption[]{Log of spectroscopic Observations }  
\label{table3}
\begin{minipage}{20cm}
\begin{tabular}{lccccll}  
\noalign{\smallskip}  

\hline
\multicolumn{7}{c}{Observing log} \\  
\hline  
Object & \multicolumn{2}{c}{Slit position} & Telescope & Exposure time & Date \\
 & $\alpha$ & $\delta$ & &  & & \\
 & (h m s) & (\degr\ \arcmin\ \arcsec) & & (sec) &  & \\  
\hline

FBP2 & 18 13 12.15 & -18 20 07.31 & 1.9 m SAAO & 900 & 24 May 2017  \\
FBP3 & 18 15 53.20 & -15 45 53.56 & 1.9 m SAAO & 1200 & 24 May 2017 \\

FBP4 & 18 19 47.14 & -12 57 01.23 & 1.9 m SAAO & 1200 & 24 May 2017  \\
FBP6 & 18 40 06.72 & -07 41 59.80 & 1.9 m SAAO & 1200 & 24 May 2017 \\

FBP7 & 18 42 15.84 & -05 00 45.00 & 1.9 m SAAO & 1200 & 24 May 2017  \\
FBP8 & 19 04 57.40 & 07 44 15.18 & 1.9 m SAAO & 1200 & 24 May 2017 \\

FBP9 & 19 21 27.80 & 15 22 54.19 & 1.9 m SAAO & 1200 & 24 May 2017 \\
FBP11 & 19 48 23.28 & 26 49 26.80 & 1.5m TUG & 3600 & 04 Jul 2017  \\

\end{tabular}
\end{minipage}
\end{table*}  

\subsection{Results}

The long-slit spectra for 8 out of our 11 PN candidates that have optical counterparts were carefully examined to confirm their nature. The spectra of 7 displayed 
nebula emission lines and the absence of a continuum except for FBP6 which may have contamination from a late-type star. Relative line fluxes and absolute 
H$\alpha$ fluxes were calculated that ranged between 1.1 and 57.5 $\times$ $10^{-15}$ erg cm$^{-2}$ s$^{-1}$ are presented in Table~4. 
The calibration errors for the signal-to-noise ratio calculations are typically less than 10\% and are 
not included in Table~4. Fig. 5 illustrates the extracted 1-D spectra of all investigated objects, including identification of the key emission lines for the two spectra at 
the top of the combined spectral plot.

Seven out of eight of the studied objects present typical PNe emission lines in the red (and in 4 cases blue) region and also typical diagnostic PNe emission line ratios. 
In particular the strong [NII]/H$\alpha$ ratios observed for most candidates eliminate H\,{\sc ii} regions as possible contaminants while the low [SII]/H$\alpha$ 
ratios ($<$0.5) eliminate Supernova remnants as contaminants. This is crucial due to the general lack of blue lines in 3 confirmed PNe due to extinction (FBP6, FBP7 and 
FBP8) preventing use in these cases of the [OIII] to H-beta ratio that is a useful PNe to H\,{\sc ii} region discriminator 
(see emission line ratio diagnostic plots in Frew \& Parker 2010) where H\,{\sc ii} regions 
with [NII]$>$0.7 are exceedingly rare. PNe diagnostic lines include [OIII] 4959, 5007\AA, [NII] and H$\alpha$ and the [SII] doublet 6716, 6731\AA, used for determining 
electron densities (see e.g. \citealt{Fr10}).  The [OIII] blue lines are only seen in 4 cases and H$\beta$ is only seen in FBP11 (the only candidate observed on the TUG 1.5m).
The observed [OIII]/H$\beta$ line ratio is high, as expected for PNe (see Table~4), though the 
S/N is poor.  The low observed fluxes of the [OIII]  4959 and 5007\AA lines  seen in the 4 objects where such lines are detected 
are due to the strong interstellar extinction. The small departures 
from the canonical 2.86 [OIII] F(5007)/F(4959) expected ratio also reflects lines of poor S/N (e.g. see tab tabulated values for FBP3). Nevertheless, the observed emission lines and their ratios confirm the PN nature of the observed 
candidates (see e.g. Canto 1981; \citealt{Acker89}; \citealt{Fr10}; \citealt{Par12b}). The high [NII]/H$\alpha$ ratios seen in 6 of our confirmed PNe would be highly unusual 
for a random sample of PNe and  also indicate that these PNe are likely of Type~I chemistry (see \citealt{Kin94}, \citealt{Par06}; \citealt{Fr10}).

The seventh object FBP11, is possibly a PNe of very high 
excitation with depleted [NII] and [SII] (see \citealt{Par06}; \citealt{Fr10}) despite its low S/N and low resolution spectrum making resolving these lines difficult.   
The lack of He~II lines in both FBP2 and FBP3 higher S/N spectra that present He~I emission imply that these two PNe are of low excitation (see \citealt{Boum03}; 
\citealt{Zha09} though extinction could also be masking any detection; \citealt{Ali16}). For three of our objects (FBP2, FBP3, FBP4) where the [SII] 6716, 6731\AA\ doublet 
is present but with low S/N ratio, a lower limit for their electron densities was estimated (see Table~4) using the STSDAS/IRAF `temden' task \citep{Sha95}. Assuming 
electron temperatures $T_c$  = $10^4$  K \citep{Zha04}, the electron densities were found to be >600 $cm^{-3}$ which is not unsurprising if they are indeed compact, 
young and so denser, less expanded PNe shells. No PN emission lines were found in the spectrum of 
FBP9. Given the presence of a continuum we consider this object's spectra is heavily contaminated by a closely adjacent star in projection. The high resolution 
UKIDSS image of this source does indicate a PN-type image lurking adjacent to this star so it is also possible that the spectrograph slit missed the true source. An IFU observation 
should help settle the matter. These PNe candidates are not visible on the finding and guide camera and placement of the``invisible" PN candidate in the narrow slit was via relative 
positioning with respect to the surrounding stars.

\begin{figure*}
\centering
\includegraphics[width=\textwidth]{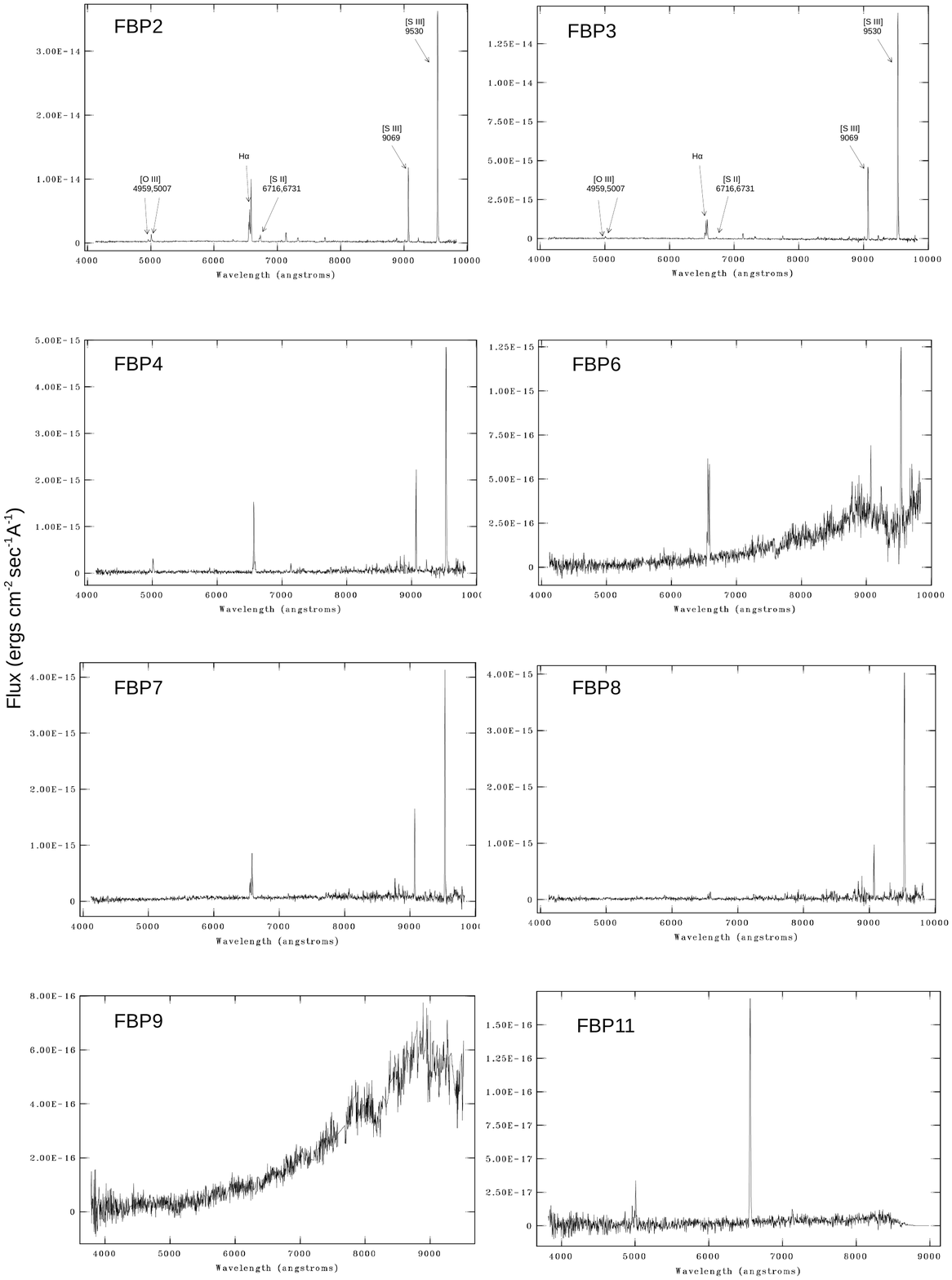}
\caption{Long--slit low resolution 1-D spectra of the observed objects. }
\label{fig6}
\end{figure*}

\begin{table*}
\caption[]{Measured relative emission line fluxes of our 7 confirmed  PNe. When the H$\beta$ line is seen as in the sole case of FBP11, the emission line ratios are calculated using the values corrected for interstellar extinction. Errors have been calculated using standard propagation.}
\label{table4}
\begin{minipage}{17cm}
\begin{tabular}{lcccccccccccc}
\hline
\noalign{\smallskip}
 & \multicolumn{3}{c}{FBP2} & \multicolumn{3}{c}{FBP3} 
& \multicolumn{3}{c}{FBP4} & \multicolumn{3}{c}{FBP6} \\ 
Line (\AA) & F\footnote{Observed fluxes normalized to F(H$\alpha$)=100 and
uncorrected for interstellar extinction.} & I\footnote{Observed fluxes normalized to F(H$\alpha$)=100 and
corrected for interstellar extinction.}  & S/N\footnote{Numbers represent the signal-to-noise ratio of the quoted fluxes.} & F & I & S/N
& F & I & S/N & F & I & S/N \\
\hline
\oxygen\ 4959 &8.4 & $-$ & 10.7 &10.1 & $-$ & 1.2 &9.4 &$-$ &3.7 &$-$ &$-$ &  $-$ \\
\oxygen\ 5007 & 26.9  & $-$  &  50.2   & 26.7  & $-$  &  5.3   & 45.3 & $-$ & 14.7 & $-$ & $-$ & $-$   \\
\nitrogen\ 6548 &48.0 & $-$ &  103.5    & 27.1 & $-$ &  16.4   &12.1 & $-$ &  3.6   &19.7 & $-$ & 5.4   \\
\ha\ 6563 &100 & $-$ &  181.2    &100 & $-$ &  50.9    &100 & $-$ &  59.4   &100 & $-$ & 26.3  \\
\nitrogen\ 6584 &196.0 & $-$ &  340.4    & 108.3 & $-$ &  51.0  &74.9 & $-$ & 17.2  & 95.1 & $-$ & 26.7  \\
\sulfur\ 6716 &8.1 & $-$ & 16.1   &3.1 & $-$ &  1.6 & 3.7 & $-$ &   1.1    & $-$ & $-$ &  $-$    \\
\sulfur\ 6731 &21.3 & $-$ & 32.1    & 7.4 & $-$ &  5.0   & 8.2 & $-$ &  2.4    & $-$ & $-$ &  $-$    \\
He I 7065 &6.4 & $-$ &  4.5    & $-$ & $-$ &  $-$   &5.9 & $-$ & 0.8  & $-$ & $-$ & $-$  \\
\ArIII\ 7136 &26.0 & $-$ & 40.4   &31.5 & $-$ &  13.0 & 24.4 & $-$ &  5.3    & $-$ & $-$ &  $-$    \\
\OII\ 7325 &18.1 & $-$ & 12.1 & 23.1 & $-$ &  2.8  & $-$ & $-$ &  $-$  & $-$ & $-$ &  $-$    \\
\ArIII\ 7752 &9.6 & $-$ & 9.1   &21.7 & $-$ &  4.2 & $-$ & $-$ & $-$   & $-$ & $-$ &  $-$    \\
P11 8862 &3.4 & $-$ & 0.3 & $-$ & $-$ &  $-$  & $-$ & $-$ &  $-$  & $-$ & $-$ &  $-$    \\
P10 9015 &8.00 & $-$ & 0.9   &26.0 & $-$ &  2.2 & $-$ & $-$ & $-$  & $-$ & $-$ &  $-$    \\
\sulfuri\ 9069 &184.5 & $-$ & 123.2 & 414.3 & $-$ &  68.0  & 234.4 & $-$ &  29.9  & 69.7 & $-$ &  2.0    \\
P9 9230 &9.1 & $-$ & 58.2  & 3.9 & $-$ &  3.2 & 25.9 & $-$ & 2.3  & $-$ & $-$ &  $-$    \\
\sulfuri\ 9530 &643.4 & $-$ & 263.0 & 1334.2 & $-$ &  136.2  & 575.7 & $-$ &  44.1  & 224.7 & $-$ &  12.2   \\
\hline

5 GHz flux (mJy) & \multicolumn{3}{c}{33.65 $\pm$ 3.3} &
\multicolumn{3}{c}{8.84 $\pm$ 1.2} & \multicolumn{3}{c}{15.37 $\pm$ 1.8} &
\multicolumn{3}{c}{9.69 $\pm$ 1.4}  \\

Absolute \ha\ flux\footnote{In units of \fluxa.} & \multicolumn{3}{c}{57.45} &
\multicolumn{3}{c}{11.78} & \multicolumn{3}{c}{8.56} &
\multicolumn{3}{c}{5.13}  \\

F(5007)/F(4959) & \multicolumn{3}{c}{3.22 $\pm$ 0.2} &
\multicolumn{3}{c}{2.65 $\pm$ 0.9} & \multicolumn{3}{c}{4.80 $\pm$ 0.9} &
\multicolumn{3}{c}{$-$}  \\

F(6583)/F(6548) & \multicolumn{3}{c}{4.09 $\pm$ 0.1} &
\multicolumn{3}{c}{3.99 $\pm$ 0.3} & \multicolumn{3}{c}{6.18 $\pm$ 0.6} &
\multicolumn{3}{c}{4.84  $\pm$ 1.2}  \\

F(6716)/F(6731) & \multicolumn{3}{c}{0.38 $\pm$ 0.03} &
\multicolumn{3}{c}{0.41 $\pm$ 0.2} & \multicolumn{3}{c}{0.45 $\pm$ 0.3} &
\multicolumn{3}{c}{$-$}  \\

\nitrogen/\ha\ & \multicolumn{3}{c}{2.44 $\pm$ 0.02} &
\multicolumn{3}{c}{1.35 $\pm$ 0.03} & \multicolumn{3}{c}{0.87 $\pm$ 0.03}
& \multicolumn{3}{c}{1.15 $pm$ 0.1 }  \\

\sulfur/\ha\ & \multicolumn{3}{c}{0.29 $\pm$ 0.01} &
\multicolumn{3}{c}{0.10 $\pm$ 0.02} & \multicolumn{3}{c}{0.12 $\pm$ 0.03}
& \multicolumn{3}{c}{$-$}  \\

n$_{[SII]}$\footnote{Electron densities in units of $cm^{-3}$. Calculated from the F(6716)/F(6731) line ratio assuming electron temperatures of $T_c = 10^4$. } & 
\multicolumn{3}{c}{> 600} &
\multicolumn{3}{c}{> 600} & \multicolumn{3}{c}{> 600}
& \multicolumn{3}{c}{$-$} \\


E$_{\rm B-V}$ (from c(H$\alpha$)) & \multicolumn{3}{c}{3.02$\pm$ 0.05} &
\multicolumn{3}{c}{3.14$\pm$ 0.07}& \multicolumn{3}{c}{3.56 $\pm$ 0.06} &  \multicolumn{3}{c}{3.58 $\pm$ 0.08} \\

Distance (kpc) & \multicolumn{3}{c}{5.5} &
\multicolumn{3}{c}{9.1} & \multicolumn{3}{c}{7.1} &
\multicolumn{3}{c}{9.2}  \\

\hline
 & \multicolumn{3}{c}{FBP7} &
\multicolumn{3}{c}{FBP8} & \multicolumn{3}{c}{FBP11}  \\ 
Line (\AA) & F & I  & S/N & F & I & S/N & F & I & S/N  \\
\hline 
\hbeta\ 4861 & $-$ & $-$ &  $-$  & $-$ & $-$ & $-$ & 2.3  & 35.1  & 2.2 \\
\oxygen\ 4959 & $-$ & $-$ &  $-$  & $-$ & $-$ & $-$ & 3.9  & 51.2  & 3.0 \\
\oxygen\ 5007 & $-$ & $-$ &  $-$  & $-$ & $-$ & $-$ & 11.9  & 141.2  & 8.2 \\
\nitrogen\ 6548 &90.0 & $-$ &  8.9   &19.1 & $-$ &  5.6   & $-$ &  $-$  &  $-$    \\
\ha\ 6563  &100 & $-$ & 10.2  &100 & $-$ & 5.7   &100 & 100  &  77.2   \\
\nitrogen\ 6584 &323.2 & $-$ & 32.9   &102.0 & $-$ &  6.8   & $-$ &  $-$  &  $-$   \\
\sulfur\ 6731 &6.4 & $-$ &  0.7   & $-$ & $-$ &  $-$  & $-$ &  $-$  &  $-$    \\
\ArIII\ 7136 &68.2 & $-$ & 0.3  & $-$ & $-$ &  $-$ & 5.1 & 3.2 &  2.6    \\
\sulfur\ 9069 &704.1 & $-$ & 33.6 & 847.6 & $-$ &  17.0 & $-$ &  $-$  &  $-$   \\
P9 9230 &111.6 & $-$ & 1.7   &59.7 & $-$ &  1.4 & $-$ & $-$ &  $-$    \\
\sulfuri\ 9530 &1914.8 & $-$ & 92.3 & 3690.5 & $-$ &  45.3  & $-$ & $-$ &  $-$    \\
\hline 

5 GHz flux (mJy) & \multicolumn{3}{c}{5.47 $\pm$ 0.1} &
\multicolumn{3}{c}{33.9 $\pm$ 3.5} & \multicolumn{3}{c}{5.23 $\pm$ 1.0}  \\

Absolute \ha\ flux & \multicolumn{3}{c}{2.32} &
\multicolumn{3}{c}{1.07} & \multicolumn{3}{c}{2.50}  \\

\oxygen/\hbeta\ & \multicolumn{3}{c}{$-$} &
\multicolumn{3}{c}{$-$} & \multicolumn{3}{c}{5.48 $\pm$
1.1}  \\

F(5007)/F(4959) & \multicolumn{3}{c}{$-$} &
\multicolumn{3}{c}{$-$} & \multicolumn{3}{c}{2.76 $\pm$ 0.6} &  \\

F(6583)/F(6548) & \multicolumn{3}{c}{3.59 $\pm$ 0.4} &
\multicolumn{3}{c}{5.33 $\pm$ 2.3} & \multicolumn{3}{c}{$-$}  \\

\nitrogen/\ha\ &\multicolumn{3}{c}{4.13 $\pm$ 0.3} &
\multicolumn{3}{c}{1.21 $\pm$ 0.3} & \multicolumn{3}{c}{$-$}\\

\sulfur/\ha\ & \multicolumn{3}{c}{0.06 $\pm$ 0.04} &
\multicolumn{3}{c}{$-$} & \multicolumn{3}{c}{$-$}  \\

c(\hbeta)\footnote{Derived by c(\hbeta) =
1/0.348$\times$log((\ha/\hbeta)$_{\rm obs}$/2.85).} & \multicolumn{3}{c}{$-$} &
\multicolumn{3}{c}{$-$} & \multicolumn{3}{c}{3.4 $\pm$ 0.2} \\

E$_{\rm B-V}$\footnote{Measured from the relation E$_{\rm B-V}$ $\approx$ 0.77c(\hbeta) \citep{Ost06}} & \multicolumn{3}{c}{$-$} &
\multicolumn{3}{c}{$-$}& \multicolumn{3}{c}{2.62 $\pm$ 0.2} \\


E$_{\rm B-V}$ (from c(H$\alpha$)) & \multicolumn{3}{c}{3.69$\pm$ 0.1} &
\multicolumn{3}{c}{4.94$\pm$ 0.1}& \multicolumn{3}{c}{3.63 $\pm$ 0.1}  \\

Distance (kpc) & \multicolumn{3}{c}{10.5} &
\multicolumn{3}{c}{5.1} & \multicolumn{3}{c}{11.2} \\
\hline
\end{tabular}
\end{minipage}
\end{table*}

\begin{table*}  
\caption[]{The measured heliocentric radial velocities of our detected PNe.}  
\label{table5}
\centering
\begin{minipage}{10 cm}

\begin{tabular}{c c c}
\hline

PN & \multicolumn{1}{c}{Heliocentric velocity (km/s)} & No. fitted lines\footnote{Number of emission lines identified by the 'rvidlines' task.} \\
\hline

FBP2 & 2.4$\pm$ 23.5 & 10  \\
FBP3 & -44.9$\pm$28.9 & 10 \\
FBP4 & 60.4$\pm$34.1 & 8  \\
FBP6 & 0.0$\pm$39.6 & 6 \\
FBP7 & 53.5$\pm$48.5 & 6  \\
FBP8 & 73.6$\pm$30.9 & 5 \\
FBP11 & -11.6$\pm$44.6 & 5 \\
\hline

\end{tabular}
\end{minipage}
\end{table*}

\section{Discussion}

We found 7 out of 8 of our observed candidate sample to be PNe with the 8th object needing re-observation due to a closely adjacent star. 
Observations are still needed for the remaining three faint PN candidates in our list with optical counterparts.  Our current results of PNe confirmation (possibly 100\% 
given the issues surrounding FBP9) gives confidence in our selection 
criteria assuming it is only extinction by intervening dust along the line of sight that prevents more optical detections. This demonstrates 
the power of radio and multi-wavelength data in the identification of new Galactic PNe that are hidden or hard to find in the optical regime. The multi-wavelength 
diagnostic tools applied to uncover PNe candidates and their subsequent confirmation demonstrates their excellent potential to detect new, highly obscured PNe.  

Although generally PNe have [NII]/H$\alpha$ line ratios < 0.6  (e.g. Frew \& Parker 2010), so called ``Type~I" PNe, first defined in \cite{Peim78} may present much stronger [NII] 
emission that can give [NII]/H$\alpha$ ratios even larger than 6 (e.g. Kerber et al. 1998; Frew, Parker \& Russeil 2006). Type~I PNe are classed as having high He and N abundances 
(see Peimbert \& Torres-Peimbert 1983) and seem to largely exhibit bipolar morphologies and high stellar temperatures (e.g. Corradi \& Schwarz 1995) and may form a distinct group 
originating from high mass progenitors. The [NII]/H$\alpha$ line ratios of Type~I PNe from the catalogue of \cite{Mac85} range from 0.72 to 2.98 having a median value of 1.38. The 
[NII]/H$\alpha$ line ratios for the sample were measured from the PNe line fluxes obtained from \cite{Web76}; \cite{All79};  \cite{All85}; \cite{Pei87}; \cite{Ack89}; \cite{Bae90}; 
\cite{Kal91}; \cite{Kin94}; \cite{Per94}; \cite{Kwi01}. The large [NII]/H$\alpha$ emission line ratios (see Table~4) from our sample indicate that at least 6 of the 7 PNe detected and 
confirmed in this work are likely of Type~I chemistry (see e.g. \citealt{Kin94}; \citealt{Ker98}; \citealt{Fr06}).  As all our PNe are compact the spectra are effectively integrated across 
the PNe so are insensitive to any positional variation in these line ratios from shocked regions. 
A higher resolution spectrum is needed for investigating FBP11 since the low resolution of our TUG spectrum 
for this object may be inadequate to resolve any [NII] emission lines if present. Our sample is highly biased in terms of selection criteria and does not reflect the chemistry 
of a volume limited  PN sample such as that of \cite{Fr06}. However, it confirms the findings of \cite{Coh11} that with a low Galactic latitude limited PN sample such as 
ours, where younger environments are prevalent, objects currently going through the PN phase will likely come from higher mass progenitors (\citealt{Peim83}; 
\citealt{Kar09}) that are also more likely to be of Type~I chemistries and have bi-polar morphologies. Higher resolution deep imaging of this sample would be useful to 
confirm this morphological assumption given the inadequacy of the current available imagery.

Our sample, though small, shows a much higher Type~I PNe rate compared to normal PNe samples and furthermore these 
usually have bipolar morphology (\citealt{Peim78}; \citealt{Peim83}). It is widely known that GLIMPSE detections are biased towards PNe 
with high mass progenitors (see \citealt{Coh11}) while the Cornish survey itself, from which this sample has been drawn, is at low Galactic latitudes which sample a younger 
environment. Any PNe found at low Galactic scale heights are likely to be young and so must derive from higher mass progenitors. A search of available spectra of previously 
identified PNe with Cornish detections using the HASH research platform, \citep{Par16} reveals that from the 30 previously known PNe that are detected in Cornish, 19 (63\%) are of 
Type~I chemistry. This indicates that the northern GLIMPSE region 
(Cornish coverage) contains more than double the high-mass progenitor PNe rate than the full GLIMPSE region (25\%, \citealt{Coh11}). 

We expect most if not all of our detected PNe to be relatively young, as highly evolved PNe have often lost their radio emission \cite{Coh11}, We also expect them 
to have more massive central stars exhibiting high stellar temperatures as is usually the case for Type~I PNe (\citealt{Tyl89}; \citealt{Coh07a}). Their compact 
nature, bright IR emission and the measured relatively high electron densities of FBP2, FBP3 and FBP4 further support their early evolutionary stage (see 
\citealt{Zha91}; \citealt{Mir10}).

The absence of detectable H$\beta$ emission line in the cases of FBP2, FBP3, FBP4, FBP6, FBP7 and FBP8 prevents us from determining the interstellar 
reddening from the Balmer decrement but following \cite{Ruf04} and using the Cornish 5 GHz radio fluxes and our measured $H\alpha$ emission line fluxes, 
assuming an electron temperature of $10^4$~K, we calculated the c(H$\alpha$) extinction of our detected PNe as: 

\smallskip
\[
c(H\alpha)=log[2.85S_v(5 GHz)/3.10 \times 10^{12} F(H\alpha)] \tag{3} \label{Eq. (3)}
\]
\medskip

Consequently the interstellar reddening was derived from the relation E$_{\rm B-V}$ $\approx$ 1.106c(H$\alpha$) \cite{Pot84} and is 
presented in Table~4. 

The compact nature of the candidates and the difficulty in determining a reliable optical angular size and $H\alpha$ flux from the current data means that we 
cannot make any reasonable distance estimations for these PNe using the H$\alpha$ surface brightness-radius ($H\alpha$-r) relation of \citep{Fr16} or the relation 
by \cite{Pier04}. However statistical distances of our detected PNe were computed using their Cornish radio data and following \cite{van95}, and these are also 
presented in Table~4 but should be adopted with caution given large uncertainties in the deconvolved radio FWHM estimates.

Although our spectral data are of relatively low resolution, the heliocentric radial velocities of our detected PNe were measured using the NOAO/IRAF rvidlines 
task and are presented in Table~5. Radial velocity measurements for a radial velocity standard star \citep{Andersen87} observed the same night as the first 6 
objects (FBP1-FBP7) indicate that their accuracy is around $\pm$~25 km/sec.

\par

\section{Conclusions}

In this work, we have not only uncovered 62 new radio selected PNe candidates and confirmed 7 as new Galactic PNe but we have also demonstrated the 
power of the use of multi-wavelength data and diagnostics in refining their selection and detection. Most of our sample of newly detected PNe are of Type~I 
chemistry indicating that the Cornish catalog contains a significantly higher rate of Type~I PNe than in the overall PNe population perhaps not surprising given the low Galactic 
latitudes of the entire sample. Their dusty environments make the use of multi-wavelength data vital for their identification. 

Further investigations should reveal the nature of the three remaining Cornish multi-wavelength selected PN candidates with optical detections, for which 
spectroscopic measurements are still missing. We confidently predict their likely PNe nature. The true status of FBP9 still needs to be confirmed though the UKIDSS data does 
suggest a true PN candidate is present.

\section*{Acknowledgements}
This research made use of data from SuperCOSMOS \ha\ Survey (AAO/UKST) and the HASH research platform. 
We are grateful to the SAAO for generous awards of telescope time for the follow-up optical spectroscopy. 
The first author thanks the University of Hong Kong for the provisions of a PhD scholarship. 
We thank TUBITAK for a partial support in using RTT150 (Russian-Turkish 1.5-m telescope in Antalya) with project number 16BRTT150-1064.







\bsp	
\label{lastpage}
\end{document}